\documentclass[aps,pra,reprint,groupedaddress]{revtex4-1}
\usepackage{amssymb}
\usepackage{amsmath}
\usepackage{array}
\usepackage{multirow,bigdelim}
\usepackage{slashbox}
\usepackage{enumitem}
\usepackage{graphicx}
\usepackage{epsfig}
\usepackage{graphics}
\usepackage{latexsym}

\begin{document}

\title{ SLOCC classification of n qubits invoking the proportional
relationships for spectrums and for standard Jordan normal forms }
\author{ Dafa Li$^{1,2}$}

\begin{abstract}
We investigate the proportional relationships for spectrums and for SJNFs
(Standard Jordan Normal Forms) of the matrices constructed from coefficient
matrices of two SLOCC (stochastic local operations and classical
communication) equivalent states of $n$ qubits. Invoking the proportional
relationships for spectrums and for SJNFs, pure states of $n$ ($\geq 4$)
qubits are partitioned into 12 groups and 34 families under SLOCC,
respectively. Specially, it is true for four qubits.
\end{abstract}


\affiliation{
$^1$Department of Mathematical Sciences, Tsinghua University,
Beijing, 100084, China\\
$^2$Center for Quantum Information Science and Technology, Tsinghua National
Laboratory for Information Science and Technology (TNList), Beijing,
100084, China}

\maketitle

\section{Introduction}

Quantum entanglement is an essential resource in quantum teleportation,
quantum cryptography, and quantum information and computation \cite{Nielsen}%
. A key task of the entanglement theory is to classify different types of
entanglement. SLOCC classification is very significant because the states in
the same SLOCC class are able to perform the same QIT-tasks \cite{Dur}\cite%
{Verstraete}. It is well known that two-qubit states were partitioned into
two SLOCC classes, three-qubit states were partitioned into six SLOCC
classes, and there are infinitely many SLOCC classes for $n$ ($\geq 4$)
qubits \cite{Dur}. It is highly desirable to partition these infinite
classes into a finite number of families according to a SLOCC invariant
criterion. In the pioneering work of Verstraete \emph{et al.} \cite%
{Verstraete}, by using their general singular value decomposition Verstraete
\emph{et al. }partitioned pure four-qubit states into nine SLOCC
inequivalent families: $G_{abcd}$, $L_{abc_{2}}$, $L_{a_{2}b_{2}}$, $%
L_{ab_{3}}$, $L_{a_{4}}$,$\ L_{a_{2}0_{3\oplus 1}}$, $L_{0_{5\oplus 3}}$, $%
L_{0_{7\oplus 1}}$, and $L_{0_{3\oplus 1}0_{3\oplus 1}}$ \cite{Verstraete}.
Since then, the extensive efforts have contributed to studying entanglement
classification of four qubits \cite{Verstraete,Miyake, Cao,
LDF07b,Chterental,Lamata,LDFQIC09,Borsten, Viehmann, Buniy, Sharma12}.

Recently, considerable efforts have been devoted to find SLOCC invariant
polynomials in the coefficients of states for classifications and measures
of entanglement of$\ n$ qubits \cite%
{Wong,Luque,Leifer,Levay,LDF07a,Osterloh09, Viehmann, Eltschka, Gour,
LDFPRA13}. It is well known that the concurrence and the 3-tangle are
invariant polynomials of degrees 2 and 4 for two and three qubits,
respectively \cite{Coffman}. Explicit and simple expresses of invariant
polynomials of degrees 2 for even $n$ qubits \cite{LDF07a}, 4 for odd $n$ ($%
\geq 4$) qubits \cite{LDF07a}, 4 for even $n$ ($\geq 4$) qubits \cite%
{LDFPRA13}, were presented.

Very recently, SLOCC invariant ranks of the coefficient matrices were
proposed for SLOCC\ classification \cite{LDFPRL12,LDFPRA12, Wang, Fan,
LDFPRA15}.

In this paper, for two SLOCC equivalent states of $n$ qubits, we show that
the matrices constructed from coefficient matrices of the two states have
proportional spectrums and proportional SJNFs. Invoking the proportional
relationships for spectrums pure states of $n$ ($\geq 4$) qubits are
partitioned into 12 groups under SLOCC, and invoking the proportional
relationships for SJNFs pure states of $n$ ($\geq 4$) qubits are partitioned
into 34 families under SLOCC. Specially, for four qubits, we obtain new
SLOCC classifications.

\section{\textbf{SLOCC classification of }$n$\textbf{\ qubits }}

\subsection{The proportional relationships for spectrums and for SJNFs}

Let $|\psi \rangle =\sum_{i=0}^{2^{n}-1}a_{i}|i\rangle $ be an $n$-qubit
pure state. It is well known that two $n$-qubit pure states $|\psi \rangle $
and $|\psi ^{\prime }\rangle $ are SLOCC equivalent if and only if there are
invertible local operators $\mathcal{A}_{i}\in GL(2,C)$, $i=1,\cdots ,n$,
such that \cite{Dur}
\begin{equation}
|\psi ^{\prime }\rangle =\mathcal{A}_{1}\otimes \mathcal{A}_{2}\otimes
\cdots \otimes \mathcal{A}_{n}|\psi \rangle .  \label{slocc}
\end{equation}

To any state $|\psi \rangle $ of $n$ qubits, we associate\ a $2^{\ell }$ by $%
2^{n-\ell }$ matrix $C_{q_{1}\cdots q_{\ell }}^{(n)}(|\psi \rangle )$ whose
entries are the coefficients $a_{0},a_{1},\cdots ,a_{2^{n}-1}$ of the state $%
|\psi \rangle $, where $q_{1},\cdots ,q_{\ell }$ are chosen as the row bits
\cite{LDFPRL12, LDFPRA12}. In \cite{LDFPRA15}, in terms of the coefficient
matrix $C_{q_{1},...,q_{i}}^{(n)}$ we constructed\ a $2^{i}$ by $2^{i}$
matrix $\Omega _{q_{1},...,q_{i}}^{(n)}$
\begin{eqnarray}
&&\Omega _{q_{1},...,q_{i}}^{(n)}(|\psi \rangle )  \notag \\
&=&C_{q_{1},...,q_{i}}^{(n)}(|\psi \rangle )\upsilon ^{\otimes
(n-i)}(C_{q_{1},...,q_{i}}^{(n)}(|\psi \rangle ))^{t},  \label{mat-1}
\end{eqnarray}%
where $\upsilon =\sqrt{-1}\sigma _{y}$ and $\sigma _{y}\ $is the Pauli
operator, and $C^{t}$ is the transpose of $C$.

From \cite{LDFPRA15}, when $q_{1}$ and $q_{2}$ are chosen as the row bits,
we can show that if $n$-qubit\ states $|\psi ^{\prime }\rangle $\ and $|\psi
\rangle $\ are SLOCC equivalent, then
\begin{eqnarray}
&&\Omega _{q_{1}q_{2}}^{(n)}(|\psi ^{\prime }\rangle )  \notag \\
&=&(\Pi _{\ell =3}^{n}\det \mathcal{A}_{q_{\ell }})(\mathcal{A}%
_{q_{1}}\otimes \mathcal{A}_{q_{2}})\Omega _{q_{1}q_{2}}^{(n)}(|\psi \rangle
)(\mathcal{A}_{q_{1}}\otimes \mathcal{A}_{q_{2}})^{t}.  \notag \\
&&  \label{g-4}
\end{eqnarray}

Let the unitary matrix
\begin{equation}
T=\frac{1}{\sqrt{2}}\left(
\begin{array}{cccc}
1 & 0 & 0 & 1 \\
0 & i & i & 0 \\
0 & -1 & 1 & 0 \\
i & 0 & 0 & -i%
\end{array}%
\right) .
\end{equation}%
Let $G_{1}=\ \ T(\mathcal{A}_{q_{1}}\otimes \mathcal{A}_{q_{2}})T^{+}$,
where $T^{+}$ is the Hermitian transpose of $T$. It is easy to check that $%
G_{1}G_{1}^{t}=(\det \mathcal{A}_{q_{1}}\ast \det \mathcal{A}_{q_{2}})I$.
Let $S_{q_{1}q_{2}}^{(n)}(\psi ^{\prime })=T\Omega _{q_{1}q_{2}}^{(n)}(|\psi
^{\prime }\rangle )T^{t}$. Then, from Eq. (\ref{g-4}) we obtain

\begin{eqnarray}
&&S_{q_{1}q_{2}}^{(n)}(\psi ^{\prime })  \notag \\
&=&(\Pi _{\ell =3}^{n}\det \mathcal{A}_{q_{\ell }})\times  \notag \\
&&T(\mathcal{A}_{q_{1}}\otimes \mathcal{A}_{q_{2}})T^{+}T\Omega
_{q_{1}q_{2}}^{(n)}(|\psi \rangle )T^{t}T^{\ast }(\mathcal{A}_{q_{1}}\otimes
\mathcal{A}_{q_{2}})^{t}T^{t}  \notag \\
&=&kG_{1}S_{q_{1}q_{2}}^{(n)}(\psi )G_{1}^{-1}  \label{ss-}
\end{eqnarray}%
where $T^{\ast }$ is a conjugate matrix, $T^{t}T^{\ast }=I$, and $k=\Pi
_{\ell =1}^{n}\det \mathcal{A}_{\ell }$. Note that $S_{q_{1}q_{2}}^{(n)}(%
\psi ^{\prime })$ and $S_{q_{1}q_{2}}^{(n)}(\psi )$ are $4$ by $4$\ matrices.

In this paper, we write the direct sum of standard Jordan blocks $%
J_{n_{1}}(\lambda _{1})$,$\cdots $, and $J_{n_{j}}(\lambda _{j})$ as $%
J_{n_{1}}(\lambda _{1})\cdots J_{n_{j}}(\lambda _{j})$. The Jordan block $%
J_{1}(a)$ is simply written as $a$. We define that the two SJNFs $%
J_{n_{1}}(\lambda _{1})\cdots J_{n_{j}}(\lambda _{j})$ and $%
J_{n_{1}}(k\lambda _{1})\cdots J_{n_{j}}(k\lambda _{j})$, where $k\neq 0$,
are proportional.

Eq. (\ref{ss-}) leads to the following theorem 1.

\textit{Theorem 1}. If the states $|\psi ^{\prime }\rangle $ and $|\psi
\rangle $ of $n$ qubits satisfy Eq. (\ref{slocc}), i.e. the state $|\psi
^{\prime }\rangle $ is SLOCC equivalent to $|\psi \rangle $, then

(1) if $S_{q_{1}q_{2}}^{(n)}(\psi )$ has the spectrum $\lambda _{1}$, $%
\cdots $, $\lambda _{4}$, then $S_{q_{1}q_{2}}^{(n)}(\psi ^{\prime })$ has
the spectrum $k\lambda _{1}$, $\cdots $, $k\lambda _{4}$, where $k=\Pi
_{\ell =1}^{n}\det \mathcal{A}_{\ell }$.

(2) if $S_{q_{1}q_{2}}^{(n)}(\psi )$ has\ the SJNF $J_{n_{1}}(\lambda
_{1})\cdots J_{n_{j}}(\lambda _{j})$, then $S_{q_{1}q_{2}}^{(n)}(\psi
^{\prime })$ has the SJNF $J_{n_{1}}(k\lambda _{1})\cdots J_{n_{j}}(k\lambda
_{j})$, where $k=\Pi _{\ell =1}^{n}\det \mathcal{A}_{\ell }$.

We give our argument as follows. Let $\Gamma =G_{1}S_{q_{1}q_{2}}^{(n)}(\psi
)G_{1}^{-1}$. Then, $S_{q_{1}q_{2}}^{(n)}(\psi ^{\prime })=k\Gamma $.
Clearly, $\Gamma $ is similar to $S_{q_{1}q_{2}}^{(n)}(\psi )$. Therefore, $%
\Gamma $ and $S_{q_{1}q_{2}}^{(n)}(\psi )$ have the same spectrum and SJNF.

(1). It is clear that if $\Gamma $ has the spectrum $\lambda _{1}$,$\cdots $%
, $\lambda _{4}$, then $S_{q_{1}q_{2}}^{(n)}(\psi ^{\prime })$ has the
spectrum $k\lambda _{1}$,$\cdots $, $k\lambda _{4}$.

(2). There is an invertible matrix $H$ such that $\Gamma =HJH^{-1}$, where
the SJNF $J=J_{n_{1}}(\lambda _{1})\cdots J_{n_{j}}(\lambda _{j})$. Then, $%
k\Gamma =HkJH^{-1}$. It is not hard to see that the SJNF\ of $kJ$ is $%
J_{n_{1}}(k\lambda _{1})\cdots J_{n_{j}}(k\lambda _{j})$.

Example 1. We have the following SLOCC\ equivalent states of four qubits: $%
L_{a_{4}}(a\neq 0)$ and $L_{a_{4}}(a=1)$ \cite{LDFQIC09}; $%
L_{a_{2}0_{3\oplus 1}}(a\neq 0)$ and $L_{a_{2}0_{3\oplus 1}}(a=1)$ \cite%
{LDFQIC09}; and $L_{ab_{3}}^{\ast }(a=0)$ and $L_{ab_{3}}(a=0)$ \cite%
{LDFPRA12}. We list the SJNFs of $S_{1,2}^{(4)}$ of the states in Table \ref%
{tab1}.

\begin{table}[tbph]
\caption{ SJNFs of SLOCC\ equivalent states }
\label{tab1}%
\begin{ruledtabular}
\begin{tabular}{cccc}
state  & $L_{a_{4}}(a\neq 0)$ & $L_{a_{4}}(a=1)$ & $k=a^{2}$
\\
SJNF & $J_{4}(a^{2})$ & $J_{4}(1)$ &  \\  \hline
State & $L_{a_{2}0_{3\oplus 1}}(a\neq 0)$ & $L_{a_{2}0_{3\oplus 1}}(a=1)$ & $%
k=a^{2}$ \\
SJNF & $J_{2}(a^{2})J_{2}(0)$ & $J_{2}(1)J_{2}(0)$ &  \\ \hline
State & $L_{ab_{3}}^{\ast }(a=0)$ & $L_{ab_{3}}(a=0)$ & $k=1$
\\
SJNF & $0b^{2}J_{2}(0)$ & $0b^{2}J_{2}(0)$ &
\end{tabular}
\end{ruledtabular}
\end{table}

Example 2. For four qubits, let $\zeta _{4}=a(|0\rangle +|15\rangle
)+b(|5\rangle +|10\rangle )+|6\rangle $, and $\zeta _{5}=b(|0\rangle
+|15\rangle )+a(|5\rangle +|10\rangle )+|6\rangle $, where $a\neq b$. The
SJNF\ of $S_{1,2}^{(4)}(\zeta _{4})$ is $J_{2}(b)aa$ while the SJNF of $%
S_{1,2}^{(4)}(\zeta _{5})$ is $J_{2}(a)bb$. So, by (2) of Theorem 1 the two
states $\zeta _{4}$ and $\zeta _{5}$ are SLOCC inequivalent.

We can rewrite $S_{q_{1}q_{2}}^{(n)}(\psi )$ as
\begin{equation}
S_{q_{1}q_{2}}^{(n)}(\psi )=[TC_{q_{1}q_{2}}^{(n)}(|\psi \rangle )]\upsilon
^{\otimes (n-2)}[TC_{q_{1}q_{2}}^{(n)}(|\psi \rangle )]^{t}.  \label{se-1}
\end{equation}

\subsection{Partition pure states of $n$ ($\geq 4$) qubits into 12 groups
and 34 families}

Theorem 1 permits a reduction of SLOCC classification of $n$ ($\geq 4$)
qubits to a classification of $4$ by $4$\ complex matrices. For $4$ by $4$\
matrices, a calculation yields 12 types of CPs (characteristic polynomials),
12 types of spectrums, and 34 types of SJNFs\ in Table \ref{tab2}. It is
easy to see that CPs and spectrums\ have the same effect for SLOCC
classification.

Note that in Table \ref{tab2}, $\sigma _{i}\neq 0$, $\sigma _{i}\neq \sigma
_{j}$ when $i\neq j$. Next, we give 12 types of CPs of $4$ by $4$\ matrices
as follows.

CP$_{1}:(\sigma -\sigma _{1})^{4}$;\ CP$_{2}:(\sigma -\sigma _{1})(\sigma
-\sigma _{2})^{3}$; CP$_{3}:(\sigma -\sigma _{1})(\sigma -\sigma
_{2})(\sigma -\sigma _{3})^{2}$; CP$_{4}:(\sigma -\sigma _{1})^{2}(\sigma
-\sigma _{2})^{2}$; CP$_{5}:\Pi _{i=1}^{4}(\sigma -\sigma _{i})$; CP$%
_{6}:\sigma (\sigma -\sigma _{1})^{3}$; CP$_{7}:\sigma (\sigma -\sigma
_{1})(\sigma -\sigma _{2})^{2}$; CP$_{8}:\sigma \Pi _{i=1}^{3}(\sigma
-\sigma _{i})$; CP$_{9}:\sigma ^{2}(\sigma -\sigma _{1})^{2}$; CP$%
_{10}:\sigma ^{2}(\sigma -\sigma _{1})(\sigma -\sigma _{2})$; CP$%
_{11}:\sigma ^{3}(\sigma -\sigma _{1})$; CP$_{12}:\sigma ^{4}$.

For each state of $n$ ($\geq 4$) qubits, the spectrum of $%
S_{q_{1}q_{2}}^{(n)}\ $must belong to one of the 12 types of the spectrums
in Table \ref{tab2}. Let the states of $n$ ($\geq 4$) qubits, for which
spectrums of $S_{q_{1}q_{2}}^{(n)}$ possess the same type in Table \ref{tab2}%
, belong to the same group. Thus, the states of $n$ ($\geq 4$) qubits are
partitioned into 12 groups. In light of Theorem 1, the states belonging to
different groups must be SLOCC inequivalent.

For each state of $n$ ($\geq 4$) qubits, the SJNF of $S_{q_{1}q_{2}}^{(n)}$
up to the order of the standard Jordan blocks must belong to one of the 34
types of the SJNFs in Table \ref{tab2}. Let the states of $n$ ($\geq 4$)
qubits with the same type of SJNFs of $S_{q_{1}q_{2}}^{(n)}$\ in Table \ref%
{tab2} up to the order of the Jordan blocks belong to the same family. Thus,
we partition the states of $n$ ($\geq 4$) qubits into 34 families. In light
of Theorem 1, the states belonging to different families must be SLOCC
inequivalent.

\begin{table}[tbph]
\caption{12 types of CPs, 12 types of spectrums, 34 types of the SJNFs for 4
by 4 matrices, and the corresponding states for four qubits.}
\label{tab2}%
\begin{ruledtabular}
\begin{tabular}{ccccc}

CP$_{i}$;spectrum & SJNF & state & SJNF & state \\ \hline

1;$\sigma _{1}\sigma _{1}\sigma _{1}\sigma _{1}$ & $J_{4}(\sigma _{1})\text{ }$
& $\tau _{1}$ & $J_{2}(\sigma _{1})J_{2}(\sigma _{1})$ &$ \eta _{1} $\\
& $J_{3}(\sigma _{1})\sigma _{1}$ & $\theta _{1}$
& $\sigma _{1}\sigma_{1}{}J_{2}(\sigma _{1})$ & $\zeta _{1} $\\
& $\sigma _{1}\sigma _{1}\sigma _{1}\sigma _{1}{}$ & $G_{1}$ &  &  \\

2;$\sigma _{1}\sigma _{2}\sigma _{2}\sigma _{2} $
& $ \sigma _{1}J_{3}(\sigma_{2}) $
& $ \theta _{2} $ & $ \sigma _{1}\sigma _{2}J_{2}(\sigma _{2}) $ & $ \zeta _{2}$
\\
 & $ \sigma _{1}\sigma _{2}\sigma _{2}\sigma _{2}{} $ & $ G_{2} $&  &  \\
3;$\sigma _{1}\sigma _{2}\sigma _{3}\sigma _{3} $ & $ \sigma _{1}\sigma
_{2}J_{2}(\sigma _{3}) $ & $ \zeta _{3} $ & $ \sigma _{1}\sigma _{2}\sigma
_{3}\sigma _{3}{} $ & $ G_{3} $\\
4;$\sigma _{1}\sigma _{1}{}\sigma _{2}\sigma _{2} $ & $ \sigma _{1}\sigma
_{1}{}\sigma _{2}\sigma _{2}{} $ & $ G_{4} $ & $ \sigma _{1}\sigma
_{1}{}J_{2}(\sigma _{2}) $ & $ \zeta _{4} $\\
 & $ J_{2}(\sigma _{1})J_{2}(\sigma _{2}) $ & $ \eta _{2} $ &  & \\
5;$\sigma _{1}\sigma _{2}\sigma _{3}\sigma _{4} $ & $ \sigma _{1}\sigma
_{2}\sigma _{3}\sigma _{4} $ & $ G_{5}$ &  &  \\
6;0$\sigma _{1}\sigma _{1}\sigma _{1} $ & $ 0J_{3}(\sigma _{1}) $ & $ \theta _{3} $ & $
0J_{2}(\sigma _{1})\sigma _{1} $ & $ \zeta _{6} $\\
& $0\sigma _{1}\sigma _{1}\sigma _{1}{} $ & $ G_{6}$ &  &  \\
7;0$\sigma _{1}\sigma _{2}\sigma _{2} $ & $ 0\sigma _{1}J_{2}(\sigma _{2}) $ & $
\zeta _{7} $ & $ 0\sigma _{1}\sigma _{2}\sigma _{2}{} $ & $ G_{7}$ \\
8;$0\sigma _{1}\sigma _{2}\sigma _{3} $ & $ 0\sigma _{1}\sigma _{2}\sigma _{3} $ & $
G_{8}$ &  &  \\
9;$00\sigma _{1}\sigma _{1} $ & $ J_{2}(0)J_{2}(\sigma _{1}) $ & $ \kappa _{1} $ & $
J_{2}(0)\sigma _{1}\sigma _{1}{} $ & $ \mu _{2} $\\
& $00J_{2}(\sigma _{1}) $ & $ \zeta _{9} $ & $ 00\sigma _{1}\sigma _{1}{} $ & $
\zeta _{8}$ \\
10;$00\sigma _{1}\sigma _{2} $ & $ J_{2}(0)\sigma _{1}\sigma _{2} $ & $ \mu _{1} $ & $
00\sigma _{1}\sigma _{2} $ & $ \zeta _{10}$ \\
11;$000\sigma _{1} $ & $ J_{3}(0)\sigma _{1} $ & $ \xi _{1} $ & $ J_{2}(0)0\sigma _{1} $ & $
\theta _{4} $\\
& $000\sigma _{1} $ & $ \zeta _{11} $&  &  \\
12;0000  &$ J_{4}(0) $ & $ L_{0_{7\oplus 1}} $ & $ J_{3}(0)0 $ & $ \xi _{2}$ \\
& $J_{2}(0)J_{2}(0) $ & $ \tau _{2} $ & $ J_{2}(0)00 $ & $ \theta _{5}$
\\
& 0000 & $ \zeta _{12}$ &  & \\

\end{tabular}
\end{ruledtabular}
\end{table}

Example 3. For the maximally entangled states $|\Psi _{2}\rangle $, $|\Psi
_{4}\rangle -|\Psi _{6}\rangle $\ of five qubits and $|\Xi _{2}\rangle $, $%
|\Xi _{4}\rangle -|\Xi _{7}\rangle $ of six qubits \cite{Osterloh06}, SJNFs
of $S_{1,2}^{(5)}$ partition $|\Psi _{2}\rangle $, $|\Psi _{4}\rangle -|\Psi
_{6}\rangle $ into three families, and SJNFs of $S_{1,2}^{(6)}$ partition $%
|\Xi _{2}\rangle $, $|\Xi _{4}\rangle -|\Xi _{7}\rangle $ into four
families. See Table \ref{tab3}.

\begin{table}[tbph]
\caption{ SJNFs of $S_{1,2}^{(5)}$ and $S_{1,2}^{(6)}$.}
\label{tab3}%
\begin{tabular}{cccccc}
\hline\hline
states & $|\Psi _{2}\rangle $ & $|\Psi _{4}\rangle $ & $|\Psi _{5}\rangle $
& $|\Psi _{6}\rangle $ &  \\
SJNFs & $\pm \frac{1}{2}00$ & $0000$ & $0000$ & $0J_{3}(0)$ &  \\
states & $|\Xi _{2}\rangle $ & $|\Xi _{4}\rangle $ & $|\Xi _{5}\rangle $ & $%
|\Xi _{6}\rangle $ & $|\Xi _{7}\rangle $ \\
SJNFs & $(\frac{1}{2})(\frac{1}{2})00$ & $0000$ & $0000$ & $00J_{2}(0)$ & $%
0J_{3}(0)$ \\ \hline\hline
\end{tabular}%
\end{table}

\section{\textbf{SLOCC classification of two, three, and four qubits }}

\subsection{SLOCC\ classification of four qubits \ }

For four qubits, invoking the fact that $T^{+}T^{\ast }=\upsilon ^{\otimes
2} $\ Eq. (\ref{se-1}) reduces to
\begin{equation}
S_{q_{1}q_{2}}^{(4)}(\psi
)=(TC_{q_{1}q_{2}}^{(4)}T^{+})(TC_{q_{1}q_{2}}^{(4)}T^{+})^{t}.
\end{equation}

From the above discussion, in light of Theorem 1 pure states of four qubits
are partitioned into 12 groups and 34 families in Table \ref{tab2}.
Furthermore, for each type of spectrums, CPs, and SJNFs in Table \ref{tab2},
we give a state in Table \ref{tab2} and the appendix for which $%
S_{1,2}^{(4)} $ has the corresponding type. For example, $S_{1,2}^{(4)}\ $of
the state $\theta _{1}$ has the spectrum $\sigma _{1}$, $\sigma _{1}$, $%
\sigma _{1}$, $\sigma _{1}$, the CP $(\sigma -\sigma _{1})^{4}$, and the
SJNF $J_{3}(\sigma _{1})\sigma _{1}$. It is plain to see that 12 groups and
34 families in Table \ref{tab2} are both complete for four qubits.

Here, we make a comparison to Verstraete et al.'s nine families. They showed
that for a complex $n$ by $n$ matrix, there are complex orthogonal matrices $%
O_{1}$ and $O_{2}$ such that $R=O_{1}R^{\prime }O_{2}$, where $R^{\prime }$
is a direct sum of blocks defined in \cite{Verstraete}. Note that the blocks
are not standard Jordan blocks. The decomposition was called a
generalization of the singular value decomposition and used to partition
pure states of four qubits into nine families \cite{Verstraete}.

Recently, Chterental and Djokovi\'{c} pointed out an error in Verstraete et
al.'s nine families by indicating that the family$\ L_{ab_{3}}$ is SLOCC\
equivalent to the subfamily $L_{abc_{2}}(a=c)$ of the family $L_{abc_{2}}$
\cite{Chterental}\cite{LDFPRA15}. Thus, the classification for the nine
families is incomplete. The need to redo this classification of four qubits
was proposed \cite{Chterental}.

\subsection{\textbf{SLOCC classification of three qubits}}

For three qubits, Eq. (\ref{se-1}) reduces to
\begin{equation}
S_{q_{1}q_{2}}^{(3)}(\psi )=[TC_{q_{1}q_{2}}^{(3)}(|\psi \rangle )]\upsilon
\lbrack TC_{q_{1}q_{2}}^{(3)}(|\psi \rangle )]^{t}.
\end{equation}

Let $\lambda
^{2}=[(c_{0}c_{7}-c_{1}c_{6})-(c_{2}c_{5}-c_{3}c_{4})]^{2}-4(c_{0}c_{3}-c_{1}c_{2})(c_{4}c_{7}-c_{5}c_{6})
$. Note that $\lambda ^{2}$ is just the $3$-tangle. The spectrum of $%
S_{1,2}^{(3)}(\psi )$ is $\pm \lambda ,0,0$. We list the SJNFs of $%
S_{1,2}^{(3)}(\psi )$ and $S_{1,3}^{(3)}(\psi )$ in the Table \ref{tab4}. In
light of Theorem 1, we can distinguish the six SLOCC classes of three
qubits. \

\begin{table}[tbp]
\caption{SLOCC classification of three qubits}
\label{tab4}%
\begin{tabular}{ccc}
\hline\hline
states & SJNF of $S_{1,2}^{(3)}(\psi )$ & SJNF of $S_{1,3}^{(3)}(\psi )$ \\
\hline
GHZ & $J_{1}(\pm \frac{1}{2})00$ & $J_{1}(\pm \frac{1}{2})00$ \\
W & $J_{3}(0)0$ & $J_{3}(0)0$ \\
A-BC & $J_{2}(0)J_{2}(0)$ & $J_{2}(0)J_{2}(0)$ \\
B-AC & $J_{2}(0)J_{2}(0)$ & $0000$ \\
C-AB & $0000$ & $J_{2}(0)J_{2}(0)$ \\
$|000\rangle $ & $0000$ & $0000$ \\ \hline\hline
\end{tabular}%
\end{table}

\subsection{\textbf{SLOCC classification of two qubits }}

For two qubits, Eq. (\ref{se-1}) reduces to

\begin{equation}
S_{1,2}^{(2)}(\psi )=[TC_{1,2}^{(2)}(|\psi \rangle )][TC_{1,2}^{(2)}(|\psi
\rangle )]^{t}.
\end{equation}%
The spectrum of $S_{1,2}^{(2)}(\psi )$\ is 0, 0, 0, $\lambda ^{\prime }$,
where $\lambda ^{\prime }=2(a_{0}a_{3}-a_{1}a_{2})$. There are two cases for
SJNFs. Case 1. For which $a_{0}a_{3}=a_{1}a_{2}$\ (it is a separate state),
the SJNF of $S_{1,2}^{(2)}(\psi )$ is $J_{2}(0)00$. Case 2. For which $%
a_{0}a_{3}\neq a_{1}a_{2}$ (it is an entangled state), the SJNF of $%
S_{1,2}^{(2)}(\psi )$ is $\lambda ^{\prime }000$. Thus, in light of Theorem
1, we can distinguish two-qubit states into two SLOCC classes.

\section{\textbf{SLOCC classification of }$n$\textbf{\ qubits under }$%
\mathcal{A}_{i}\in SL(2,C)$}

SLOCC classification under $A_{i}\in SL(2,C)$ or the classification under
determinant one SLOCC operations was discussed in previous articles \cite%
{Verstraete}\cite{Luque}. Note that under $\mathcal{A}_{i}\in SL(2,C)$,\ $%
G_{1}\in SO(4,C)$ and Eq. (\ref{ss-}) reduces to

\begin{equation}
S_{q_{1}q_{2}}^{(n)}(\psi ^{\prime })=G_{1}S_{q_{1}q_{2}}^{(n)}(\psi
)G_{1}^{-1}.  \label{sl-1}
\end{equation}%
\ Thus, Eq. (\ref{sl-1}) leads to the following theorem.

\textit{Theorem 2.} If the states $|\psi ^{\prime }\rangle $ and $|\psi
\rangle $ of $n$ qubits are SLOCC equivalent under $\mathcal{A}_{i}\in
SL(2,C)$, then $S_{q_{1}q_{2}}^{(n)}(\psi ^{\prime })$ is\ orthogonally
similar to $S_{q_{1}q_{2}}^{(n)}(\psi )$. The similarity implies that $%
S_{q_{1}q_{2}}^{(n)}(\psi ^{\prime })$ and $S_{q_{1}q_{2}}^{(n)}(\psi )$
have the same CP, spectrum, and SJNF up to the order of the standard Jordan
blocks.

Example 4. $L_{ab_{3}}^{\ast }(a=0)$ is SLOCC equivalent to $L_{ab_{3}}(a=0)$
under $A_{i}\in SL(2,C)$ \cite{LDFPRA12}. The SJNFs of $S_{1,2}^{(4)}$ are
both $0b^{2}J_{2}(0)$.

Restated in the contrapositive the theorem reads: If two matrices $%
S_{q_{1}q_{2}}^{(n)}$ associated with two n-qubit pure states differ in
their CPs, spectrums, or SJNFs, then the two states are SLOCC\ inequivalent
under $\mathcal{A}_{i}\in SL(2,C)$. From Example 2, by Theorem 2 the two
states $\zeta _{4}$ and $\zeta _{5}$ are SLOCC inequivalent under $A_{i}\in
SL(2;C)$ because SJNFs of $S_{1,2}^{(4)}(\zeta _{4})$ and $%
S_{1,2}^{(4)}(\zeta _{5})$\ are different.

Note that a SLOCC\ equivalent class may include infinite SLOCC\ equivalent
classes under $\mathcal{A}_{i}\in SL(2,C)$.

\section{Conclusion}

In Theorem 1, we demonstrate that for two SLOCC equivalent states, the
spectrums and SJNFs of the matrices $S_{q_{1}q_{2}}^{(n)}$ have proportional
relationships. Invoking the proportional relationships, we partition pure
states of $n$ ($\geq 4$) qubits into 12 groups and 34 families under SLOCC,
respectively.

In Theorem 2, we deduce that for two equivalent states under determinant one
SLOCC operations, the spectrums, CPs, SJNFs of $S_{q_{1}q_{2}}^{(n)}$ are
invariant. The invariance can be used for SLOCC classification of $n$ qubits
under determinant one SLOCC operations.

To make a comparison, we list the differences between Theorems 1 and 2 in
Table \ref{tab5}.

It is known that SJNF is used to solve a system of linear differential
equations. The \ classification of SJNFs under SLOCC in this paper\ seems to
be useful for classifying linear differential systems.
\begin{table}[tbp]
\caption{Comparison between Theorems 1 and 2}
\label{tab5}%
\begin{tabular}{ccc}
\hline\hline
& Theorem 1 & Theorem 2 \\ \hline
spect. $\psi $ & $\lambda _{1}$, $\cdots $, $\lambda _{4}$ & $\lambda _{1}$,
$\cdots $, $\lambda _{4}$ \\
spect. $\psi ^{\prime }$ & $k\lambda _{1}$, $\cdots $, $k\lambda _{4}$ & $%
\lambda _{1}$, $\cdots $, $\lambda _{4}$ \\ \hline
SJNF $\psi $ & $J_{\ell _{1}}(\lambda _{1})\cdots J_{\ell _{j}}(\lambda
_{j}) $ & $J_{\ell _{1}}(\lambda _{1})\cdots J_{\ell _{j}}(\lambda _{j})$ \\
SJNF $\psi ^{\prime }$ & $J_{\ell _{1}}(k\lambda _{1})\cdots J_{\ell
_{j}}(k\lambda _{j})$ & $J_{\ell _{1}}(\lambda _{1})\cdots J_{\ell
_{j}}(\lambda _{j})$ \\ \hline\hline
\end{tabular}%
\end{table}

Acknowledgement---This work was supported by NSFC (Grant No. 10875061) and
Tsinghua National Laboratory for Information Science and Technology.

\section{Appendix Corresponding states of four qubits\ }

Using $G_{abcd}$ we obtain the following 8 states.

$G_{1}=G_{abcd}(a=b=c=d\neq 0)$; (we will omit $G_{abcd}$ next);

$G_{2}:abcd\neq $ $0;b=c=d$ but $a\neq b$;

$G_{3}:abcd\neq $ $0$, two of $a,b,c$, and $d$ are equal while the other two
are not equal;

$G_{4}:abcd\neq 0$, $a,b$, $c$, and $d$ consists of two pairs of equal
numbers;

$G_{5}:abcd\neq $ $0,$ $a,b$, $c$, and $d$ are distinct;

$G_{6}:$ only one of $a,b$, $c$, and $d$ is zero and other three are equal;

$G_{7}:$ only one of $a,b$, $c$, and $d$ is zero and only two of them are
equal;

$G_{8}:$ only one of $a,b$, $c$, and $d$ is zero and the other three are
distinct.

Using $L_{abc_{2}}$ we obtain the following 11 states.

$\zeta _{1}=L_{abc_{2}}($ $a=b=c\neq $ $0);$

$\zeta _{2}=L_{abc_{2}}(abc\neq 0$ and one of $a$ and $b$ equals c$)$;

$\zeta _{3}=L_{abc_{2}}(abc\neq 0$ and $a,b,c$ are distinct.$)$;

$\zeta _{4}=a(|0\rangle +|15\rangle )+b(|5\rangle +|10\rangle )+|6\rangle $,
where $a\neq b$;

$\zeta _{6}=L_{abc_{2}}($ only one of $a$ and $b$ is zero while the other is
equal to $c$.$)$;

$\zeta _{7}=L_{abc_{2}}(c\neq 0$ and only one of $a$ and $b$ is zero while
the other is not equal to $c$.$)$;

$\zeta _{8}=L_{abc_{2}}(c=0$ and $a=b\neq 0)$; $\zeta _{9}=L_{abc_{2}}(c\neq
0$ and $a=b=0)$; $\zeta _{10}=L_{abc_{2}}(c=0$ while $ab\neq 0$ and $a\neq
b) $; $\zeta _{11}=L_{abc_{2}}(c=0$ while only one of $a $ and $b$ is zero$)$%
; $\zeta _{12}=L_{abc_{2}}(a=b=c=0)$.

Using $L_{a_{2}b_{2}}$ we obtain the following two states.

$\eta _{1}=L_{a_{2}b_{2}}(a=b\neq 0)$; $\eta _{2}=L_{a_{2}b_{2}}(ab\neq 0$
and $a\neq b)$.

Let $L_{ab_{3}}^{\prime }=b(|0\rangle +|15\rangle )+\frac{b+a}{2}(|5\rangle
+|10\rangle )+\frac{b-a}{2}(|6\rangle +|9\rangle )+\frac{i}{\sqrt{2}}%
(|1\rangle +|2\rangle -|7\rangle -|11\rangle )$. Using $L_{ab_{3}}^{\prime }$
we obtain the following five states.

$\theta _{1}=L_{ab_{3}}^{\prime }(a=b\neq 0)$;

$\theta _{2}=L_{ab_{3}}^{\prime }(ab\neq 0$ and $a\neq b)$;

$\theta _{3}=a(|0\rangle +|15\rangle )+\frac{a}{2}(|5\rangle +|10\rangle
+|6\rangle +|9\rangle )+\frac{i}{\sqrt{2}}(|1\rangle +|2\rangle -|7\rangle
-|11\rangle )$ (obtained from $L_{ab_{3}}^{\prime }(a=0$ but $b\neq 0)$);

$\theta _{4}=L_{ab_{3}}^{\prime }(b=0$ but $a\neq 0)$; $\theta
_{5}=L_{ab_{3}}^{\prime }(a=b=0)$.

Using $L_{a_{4}}$ we obtain the following two states.

$\tau _{1}=L_{a_{4}}(a\neq 0)$; $\tau _{2}=L_{a_{4}}(a=0)$.

Using $L_{a_{2}0_{3\oplus 1}}$ we obtain the following one state.

$\kappa _{1}=L_{a_{2}0_{3\oplus 1}}(a\neq 0)$.

Let $L_{ab0_{3\oplus 1}}=\frac{a+b}{2}(|0\rangle +|15\rangle )+\frac{a-b}{2}%
(|3\rangle +|12\rangle )+|5\rangle +|6\rangle $. Using $L_{ab0_{3\oplus 1}}$
we obtain the following two states.

$\mu _{1}=L_{ab0_{3\oplus 1}}(ab\neq 0$ and $a\neq b)$; $\mu
_{2}=L_{ab0_{3\oplus 1}}(a=b\neq 0)$ ;

Let $\xi =\frac{a}{2}(|0\rangle +|3\rangle +|12\rangle +|15\rangle
)+i|1\rangle -i|13\rangle +|10\rangle $. Using $\xi $ we obtain the
following two states.

$\xi _{1}=\xi (a\neq 0)$; $\xi _{2}=\xi (a=0)$.

\end{document}